\documentclass[
    onecolumn,
    preprintnumbers,
    amsmath,
    amssymb,
    prd,
    showpacs
]{revtex4}
\usepackage{graphicx}%
\usepackage{amsmath}
\usepackage[normalem,normalbf]{ulem}
\usepackage{amssymb}
\usepackage{bm}
\usepackage{enumerate}

\newcommand{\nn}{\nonumber}
\newcommand{\dsp}{\displaystyle}
\begin{document}
\title{Stationary vacuum hyper-cylindrical solution in 4+1 dimensions}

\author{Jungjai Lee}
\email{jjlee@daejin.ac.kr}
\affiliation{ Department of Physics, Daejin University,
Pocheon, 487-711, Korea.
}%
\author{Hyeong-Chan Kim}
\email{hckim@phya.yonsei.ac.kr}
\affiliation{Department of Physics, Yonsei University,
Seoul 120-749, Republic of Korea.
}%
\date{\today}%
\bigskip
\begin{abstract}
\bigskip

We find a $4+1$ dimensional stationary vacuum hyper-cylindrical
solution solution which is spherically symmetric in $3-$dimensions
and invariant under the translation along the fifth coordinate.
The solution is characterized by three parameters, mass, tension,
and conserved momentum along the fifth coordinate. The metric is
locally equivalent to the known static solution. We briefly
discuss its physical properties.

\end{abstract}
\maketitle

\section{Introduction}
Recently, a generalized static hyper-cylindrical solution with arbitrary
tension $\tau$ and mass density $M$ in (4+1)-dimensions has been found
by Lee~\cite{lee}. In fact, the solution was mentioned by other authors
in eighties in a different context of Kluza-Klein gauge
interaction~\cite{gross,davidson}. These higher dimensional solutions
have become objects of serious consideration in physics such as the
string theory~\cite{GL} and brane cosmology~\cite{randal1,randal2,hawking}.

In studying these higher dimensional black string solutions, their
stability becomes an interesting topic~\cite{GL2,GM}. Various
black strings were shown to be unstable under small
perturbations~\cite{GL2,KL,hirayama}. The possibility of the
unstable black string finally fragmenting into black holes was
mentioned~\cite{choptuik}. However, Horowitz and
Maeda~\cite{horowitz} argued that event horizons could not pinch
off.

In Ref.~\cite{lee}, the asymptotic properties of the metric up to
$O(1/r)$ in asymptotic region was investigated and the static
metric was studied. In this stage, it is interesting to search for
a stationary solution with conserved momentum in the fifth
coordinate. In fact, if there is an object with momentum, the
velocity frame dragging effects do seem to appear in general
relativity. The velocity frame dragging is not well known and
controversial. In addition, its effects may alter local or global
geometry. In this sense, it needs to study the effect of a
conserved momentum on the geometry of stationary spacetime.

In this work, the static hyper-cylindrical solution~\cite{lee} is
briefly reviewed in Sec. II. A general class of hyper-cylindrical
solutions with arbitrary tension and conserved momentum is presented in
Sec. III. The local equivalence and global difference to the static
hyper-cylindrical solution are analyzed in Sec. IV. Physical properties
of the solution is briefly discussed in Sec.~V. In Sec. VI, we discuss
the geometric properties of the horizon area, and the global difference
of the metric.

\section{Brief review on the static hyper-cylindrical solution}
After the reappearance of the static hyper-cylindrical
solution~\cite{lee}, its geometric properties are being
studied~\cite{kang} especially in relation to the tension of the string.
The metric of the static spherically symmetric vacuum hyper-cylindrical
solution in (4+1) dimensions in Ref.~\cite{lee} is
\begin{eqnarray} \label{lee}
ds^2 &=&-\left|\frac{1-K/\rho}{1+K/\rho}\right|^{
    \frac{2(\chi+1/\sqrt{3})}{\sqrt{1+\chi^2}}}
    dt^2+
    \left|\frac{1-K/\rho}{1+K/\rho}\right|^{
    \frac{2(-\chi+1/\sqrt{3})}{\sqrt{1+\chi^2}}}
    dz^2 \\
&& +\left|1-\frac{K^2}{\rho^2}\right|^2
    \left|\frac{1-K/\rho}{1+K/\rho}\right|^{-
    \frac{4}{\sqrt{3}\sqrt{1+\chi^2}}}
    \left(d\rho^2+
    \rho^2 d\theta^2+ \rho^2\sin^2\theta d\phi^2\right), \nn
\end{eqnarray}
where
\begin{eqnarray} \label{K:}
K =\frac{G_5}{2\sqrt{3}}
    (M+\tau)\sqrt{1+\chi^2}, ~~~~
    \chi = \sqrt{3}\frac{M-\tau}{M+\tau}=\sqrt{3}\frac{1-a}{1+a},
\end{eqnarray}
with $a=\tau /M$. Here $G_5$ is the five dimensional gravitational
constant.
Let us briefly summarize the properties of the metric.
\begin{itemize}

\item  $\rho =K$ is a naked curvature singularity for
$\displaystyle |\chi| \neq \frac 1{\sqrt{3}}$.

\item The singularity is hidden by the horizon for
$\dsp\chi=\frac{1}{\sqrt{3}}$, which is the Schwarzschild black string
solution.

\item The strong energy condition restricts $a\leq 2~(\chi
\geq -1/\sqrt{3})$, which gives physical range of parameters $0\leq
a\leq 2~(-1/\sqrt{3} < \chi \leq \sqrt{3})$.

\item Solution with $\chi=1/\sqrt{3}$ and $\chi=-1/\sqrt{3}$
denote the Schwarzschild black string solution and the Kluza-Klein
bubble solution~\cite{harmark}, respectively.
\end{itemize}
Not all physical properties of the solution is yet understood and is
under investigation~\cite{kang}.  This solution was found by noting that
two independent asymptotic quantities are needed to characterize the
static vacuum hyper-cylindrical solution in
$5-$dimension~\cite{harmark2}, \cite{kol2}.

Let us generalize the discussion on the asymptotic quantities used
in Ref.~\cite{lee} to a stationary case in $(4+1)$ dimensional
spacetime with the coordinates $x^0=t, ~x^i(i=1,2,3)$, and
$x^4=z$. For the weak gravitational field produced by stationary
$z-$independent source, the linearized Einstein equation in
harmonic coordinates is
\begin{eqnarray} \label{LinEq}
\partial_i \partial^i h_{\mu\nu}= -16\pi G_5 \bar
T_{\mu\nu},
\end{eqnarray}
where
\begin{eqnarray} \label{h}
h_{\mu\nu}&=&g_{\mu\nu}-\eta_{\mu\nu}~~ (|h_{\mu\nu}|\ll
1), \nn\\
\bar T_{\mu\nu} &\equiv & T_{\mu
    \nu}-\frac{1}{3}\eta_{\mu\nu}T . \nn
\end{eqnarray}
Using the Green's function for the three dimensional Laplacian,
the solution of Eq.~(\ref{LinEq}) can be written by
\begin{eqnarray} \label{h:Sol}
h_{\mu\nu} (x) = 4G_5 \int d^3 y \frac{\bar T_{\mu\nu}(y)}
    {|\bf x-y|}. \nn
\end{eqnarray}
The leading term of $h_{\mu \nu}(x)$ are then calculated to be
\begin{eqnarray} \label{hs}
h_{00} &\simeq & \frac{4G_5}{r} \int d^3 y
\left(\frac{2}{3}T_{00}+\frac{1}{3}T_{44}\right), \nn \\
h_{04} &\simeq & \frac{4G_5}{r} \int d^3 y T_{04},\nn \\
h_{ij} &\simeq & \frac{4G_5\delta_{ij}}{3r} \int d^3 y
\left(T_{00}-T_{44}\right), \nn\\
h_{44} &\simeq & \frac{4G_5}{r} \int d^3 y
\left(\frac{1}{3}T_{00}+\frac{2}{3}T_{44}\right), \nn
\end{eqnarray}
where $r=\sqrt{\sum_{i=1}^3 x_i^2}$.

For the stationary case with $T^{04}\neq 0$, the leading
corrections to the metric far away from the source, up to the
order of $1/r$, can be seen to be characterized by the three
quantities
\begin{eqnarray} \label{quantities}
M \equiv \int d^3 x\, T_{00}, ~~~\tau \equiv \int d^3 x \,T_{44},
~~~P\equiv\int d^3 x\, T_{04} ,
\end{eqnarray}
where $M$ is the mass per unit length, $\tau$ is the tension, and
$P$ is the conserved momentum along the $z-$direction of the
source. For later convenience, we use the parameter $j$,
\begin{eqnarray}
j= \frac{2 P}{M-\tau},
\end{eqnarray}
in prefer of $P$. Then, up to the order of $1/r$,
\begin{eqnarray} \label{h:asym}
h_{00}(x) \simeq \frac{4G_5}{3} \frac{2M -\tau}{r}, ~~
 h_{ij}(x) \simeq \frac{4G_5}{3} \delta_{ij}
    \frac{M +\tau}{r},
    ~~
 h_{44}(x) \simeq \frac{4G_5}{3}
    \frac{M -2\tau}{r},~~
 h_{04}(x) \simeq \frac{2G_5(M-\tau) j}{r}.
\end{eqnarray}
If the coordinate $z$ is periodic with $0\leq z < L$, the four
dimensional gravitational constant $G_4$ is given by $G_4= G_5/L$ and
the total mass of the source is $M L$. Now, it is interesting to see if
the asymptotic quantity $j$ develops a new kinds of hyper-cylindrical
solution.

\section{A stationary hyper-cylindrical solution with momentum in
$z$-direction}

Let us search for the metric whose source has the tension $\tau$
and the conserved momentum $P$. Then we know, from
Eq.~(\ref{h:asym}), that the leading corrections of the metric far
away from the source are
\begin{eqnarray} \label{h:asym:gen}
h_{00}= \frac{4G_5M(2-a)}{3\rho},
~~h_{ij}=\delta_{ij}\frac{4G_5M(1+a)}{3\rho}, ~~h_{44}=
\frac{4G_5M(1-2a)}{3\rho}, ~~h_{04}=\frac{2G_5M(1-a)j}{\rho} .
\end{eqnarray}
The asymptotic form of the metric is
\begin{eqnarray} \label{ds2:asym}
ds^2 &\approx&
-\left(1-\frac{4G_5M(2-a)}{3\rho}\right)dt^2+\left(
    1+\frac{4G_5M(1+a)}{3\rho} \right)\left(d\rho^2+
    \rho^2 d\theta^2+\rho^2\sin^2\theta d\phi^2\right)\\&&+
    \left(1+\frac{4G_5M(1-2a)}{3\rho}\right)dz^2
    +\frac{4G_5 M(1-a)j}{\rho}dt dz. \nn
\end{eqnarray}
We have to find the solutions to the vacuum Einstein field
equations which reduce to the asymptotic form of
Eq.~(\ref{ds2:asym}) at large $\rho$. We start with the ansatz
\begin{eqnarray} \label{metric:ansatz}
ds^2 =-F(\rho)[dt-X(\rho) dz]^2
+G(\rho)\left(d\rho^2+\rho^2d\theta^2+
    \rho^2\sin^2\theta d\phi^2\right)+
    H(\rho) dz^2,
\end{eqnarray}
and substitute this form of metric into the vacuum Einstein field
equation to derive differential equations for the four functions
$F(\rho)=e^{2U}$, $G(\rho)=e^{2V}$, $H(\rho)=e^{2W}$, and
$X(\rho)$:
\begin{eqnarray} \label{eq}
&&U''+(U'+V'+W'+\frac{2}{\rho})U'+\frac{(X')^2}{2}e^{2(U-W)}=0,
    \\
&&U''+2V''+W''+(U')^2+(W')^2-(U'+W'-\frac{2}{\rho})V'-
    \frac{(X')^2}{2}e^{2(U-W)}=0, \nn \\
&&V''+(U'+V'+W'+\frac{2}{\rho})V'+\frac{U'+V'+W'}{\rho}=0,
 \nn \\
&&W''+(U'+V'+W'+\frac{2}{\rho})W'-\frac{(X')^2}{2}e^{2(U-W)}=0,
     \nn \\
&&X''+ (U'+V'+W'+\frac{2}{\rho})X'+2(U'-W')X'=0 . \nn
\end{eqnarray}
The above equations can be solved for $U+V+W$, $U+W$, and
$e^{2(U-W)}X'$. After solving these, by summing the first two
equations of~Eq.~(\ref{eq}), we obtain
\begin{eqnarray} \label{diffeq}
D'+D^2= \frac{3K^2(5-b^2)}{(\rho^2-K^2)^2} ,
\end{eqnarray}
where $\displaystyle
D=U'-W'+\frac{U'+V'+W'}{2}+\frac{1}{\rho}$ and $b$ is a
free parameter. A specific solution to Eq.~(\ref{diffeq})
can be found to be $D_0=(\sqrt{16-3b^2} +
\rho)/(\rho^2-K^2)$. Then, by setting $D= D_0+1/f$, we get
a linear differential equation for $f$
\begin{eqnarray} \label{f:eq}
f'-2 D_0 f=1 ,
\end{eqnarray}
whose general solution can be obtained easily.

After changing the parameter $\dsp
b=\frac{4}{\sqrt{3}\sqrt{1+\chi^2}}$, we have found the
solution of the Einstein equation:
\begin{eqnarray} \label{Sol}
ds^2_{new} &=&-\frac{1}{1-q^2}\left|\frac{1+K/\rho}{1 -
K/\rho}\right|^{-\frac{2(\chi+1/\sqrt{3})}
    {\sqrt{1+\chi^2}}}
    \left(1- q^2
    \left|\frac{1+K/\rho}{1 -
K/\rho}\right|^{\frac{4\chi}{\sqrt{1+\chi^2}}}\right)
    \left(dt + q
    \frac{1-\left|\frac{1+K/\rho}{1 -
K/\rho}\right|^{\frac{4\chi}{\sqrt{1+\chi^2}}}}{
    1-q^2\left|\frac{1+K/\rho}{1 -
K/\rho}\right|^{\frac{4\chi}{\sqrt{1+\chi^2}}}}dz \right)^2\\
&+&\left|1-\frac{K^2}{\rho^2}\right|^{2}
    \left|\frac{1+K/\rho}{1-K/\rho}\right|^{\frac{4}{\sqrt{3}
        \sqrt{1+\chi^2}}} \left(d\rho^2+
    \rho^2 d\theta^2+\rho^2\sin^2\theta d\phi^2\right)+
    \frac{(1-q^2)\left|\frac{1+K/\rho}{1 -
K/\rho}\right|^{\frac{2(\chi-1/\sqrt{3})}
    {\sqrt{1+\chi^2}}}}{
    1-q^2\left|\frac{1+K/\rho}{1 -
K/\rho}\right|^{\frac{4\chi}
    {\sqrt{1+\chi^2}}}} dz^2 \nn ,
\end{eqnarray}
where the parameters $K$, $\chi$, and $q$ are related with
the asymptotic parameters $M$, $a$, and $j$ by
\begin{eqnarray} \label{bK}
K &=& \frac{G_5}{2\sqrt{3}}
    (1+a)M\sqrt{1+\chi^2}\, , \nn\\
\chi &\equiv&\chi(a,q) =
\sqrt{3}\frac{1-a}{1+a}\frac{1-q^2}{1+q^2}
        ,\\
q^2&-&2j^{-1}q +1 =0 . \nn
\end{eqnarray}
The solution is symmetric with respect to the following
transformation of parameters:
\begin{eqnarray} \label{symmetries}
q, ~ \chi &\longrightarrow & q^{-1}, ~ -\chi . \nn
\end{eqnarray}
With this, we can restrict $q$ to the finite region $-1\leq q \leq
1$. In summary, the ranges of free parameters are given by
\begin{eqnarray} \label{parameters}
-\sqrt{3} < \chi\leq \sqrt{3}, ~~~-\infty <K <\infty,
~~~~-1\leq q \leq 1 .
\end{eqnarray}
Because of the last equality of Eq.~(\ref{bK}) and the restriction on
the range of $q$, the relation between $j$ and $q$ becomes bijective.
Therefore, we use $q$ instead of $j$ as a physical parameter which
determines the conserved momentum. We have a unique stationary
hyper-cylindrical solution described by $(\chi, K, q)$ for a given
asymptotic condition, $(M,\tau,P)$.

\section{Local equivalence to the static hyper-cylindrical solution}
The curvature square of the metric~(\ref{Sol}) is
\begin{eqnarray} \label{curv:square}
R_{\mu\nu\rho\sigma}R^{\mu\nu\rho\sigma} &=&
    \frac{192K^2}{\rho^6\left(1-K^2/\rho^2\right)^8}
    \left|\frac{1-K/\rho}{1+K/\rho}\right|^{\frac{8}{\sqrt{3}
        \sqrt{1+\chi^2}}}
    \left[\frac{K^4}{\rho^4}+1 \right.\\
&&\left.-\frac{4\sqrt{3}}{\sqrt{1+\chi^2}}
 \left(1-\frac{4}{9(1+\chi^2)}\right)
    \frac{K}{\rho}\left(\frac{K^2}{\rho^2}+1\right)
  +4\left(1+\frac{4}{3
        (1+\chi^2)}-\frac{8}{9(1+\chi^2)^2}\right)
        \frac{K^2}{\rho^2}\right]\nn .
\end{eqnarray}
Only at $|\chi|=1/\sqrt{3}$, the terms inside the square bracket is
factorized to $\left(1-K/\rho\right)^4$. Then, in addition, the exponent
of the absolute value becomes 4. For $|\chi|= 1/\sqrt{3}$, the curvature
square at $\rho=K$ is finite, $3/(64K^4)$. However, the curvature square
diverges at $\rho = K$ for $|\chi|\neq 1/\sqrt{3}$.  The point $\rho =0$
is a regular point with zero curvature. In fact, if one perform a
coordinate transform $K/\rho \rightarrow \bar \rho/K$, one gets the same
form of metric as Eq.~(\ref{Sol}), which implies that the region with
$\rho\sim 0$ is also asymptotically flat. Note that this curvature
square is independent of $q$. In addition, one can show that many
curvature invariants of lower powers of $R_{\alpha\beta \mu\nu}$,
$R_{\alpha\beta\mu\nu;\rho}$, and $R_{\alpha\beta\mu\nu;\rho\sigma}$ are
independent of $q$ by using computer program such as Mathematica. At
first glance, this seems to be a signal that the present solution is
just a gauge artifact of the static solution~(\ref{lee}).

If two metrics are equivalent to each other, it implies that both
the local and the global properties of the metrics are the same.
The local equivalence of two metrics can be tested by comparing
the Riemann curvature tensors and its covariant derivatives at
each point of the spacetime. The global equivalence can be checked
by analyzing the global structures of the spacetime.

For two equivalent metrics, there exists a general coordinate
transformation which relates the two. Because it is not easy to find the
explicit form of the transformation, we develop another method in the
present case. Two equivalent metrics should be the same up to a Lorentz
transformation at each event. At the present stationary metric, to check
the local equivalence to the static hyper-cylindrical metric, we
investigate the $q-$dependence of the local geometric quantities. Once
every local properties such as the Riemann tensors and their derivatives
at an event are the same, the two metrics are equivalent at the event.
Therefore, we examine the components of the Riemann tensor at a
spacetime event ${\cal P} =(t,\rho,\theta,\phi,z)$ in a locally
orthonormal frame of reference, given by the 1-form basis:
\begin{eqnarray} \label{coor}
\omega^0 = \left[F^{1/2}(dt - X dz)\right]_{\cal P}, ~~ \omega^1 =
\left[G^{1/2} d\rho\right]_{\cal P}, ~~\omega^2 = \left[G^{1/2}
\rho d\theta\right]_{\cal P}, ~~ \omega^3 = \left[G^{1/2} \rho
\sin\theta d\phi\right]_{\cal P}, ~~~\omega^4 = \left[H^{1/2}
dz\right]_{\cal P}.
\end{eqnarray}
Since the locally orthonormal frame of reference is unique up to
Lorentz transformation at an event ${\cal P}$, the equivalence of
the metric~(\ref{Sol}) and (\ref{lee}) at ${\cal P}$ can be
checked. In Appendix A, by considering the $\rho-$dependent
Lorentz boost along the $z$-direction,
\begin{eqnarray} \label{boost}
\left(\begin{tabular}{c}
    ${x'}^0$\\ ${x'}^4$ \\\end{tabular}\right)
    &=&\left(\begin{tabular}{c}
    $\cosh \theta~~ \sinh\theta$\\ $\sinh\theta~~
        \cosh\theta$ \\ \end{tabular}\right)
\left(\begin{tabular}{c}
    $x^0$\\ $x^4$ \\\end{tabular}\right),
\end{eqnarray}
where $x^i$ is a locally defined coordinate system in the orthonormal
frame, we show that the components of the boosted Riemann tensor at
${\cal P}$ for the new metric~(\ref{Sol}) become independent of $q$ by
choosing an appropriate boost parameter $\theta$. In this sense, the new
stationary solution is locally equivalent to the static
hyper-cylindrical solution~(\ref{lee}).

However, an interesting inspection happens at $\rho=\rho_e$ in
Eq.~(\ref{rhoe}) where $\tanh\theta=1$. Around this point, the
coordinate transformation is given by
\begin{eqnarray} \label{theta:cond}
\tanh \theta =\left[ q\left|\frac{1+K/\rho}{1-K/\rho}
            \right|^{\frac{2\chi}{\sqrt{1+\chi^2}}}\right]^{
                \mbox{sign}(\rho-\rho_e)} .
\end{eqnarray}
Therefore, the coordinate transformation is continuous at
$\rho_e$.
Its first derivative,
\begin{eqnarray} \label{theta':cond}
\frac{\partial}{\partial \rho}\tanh \theta = -
\frac{4\chi}{\sqrt{1+\chi^2}}
\frac{\mbox{sign}(\rho-\rho_e)}{\left|\frac{1+K/\rho_e}{
        1-K/\rho_e} \right|(\rho_e-K)^2 },
\end{eqnarray}
is, however, discontinuous at $\rho_e$.
Let us search for the effect of this discontinuity to the Riemann
tensor. The effect on the components of the Riemann tensor appears
in a form of combination:
\begin{eqnarray} \label{R..}
&&(\cosh^2\theta+\sinh^2\theta)R_q-2\cosh \theta
    \sinh\theta \,r_q = -\mbox{sign}(\rho-\rho_e)\sqrt{R_q^2-r_q^2}\,
    ,
\end{eqnarray}
where $R_q$ and $r_q$ are given in Appendix A. The right hand side
of Eq.~(\ref{R..}) is discontinuous at $\rho_e$ for all $(q, \chi
>0 )$ except for the pair given by the relation
\begin{eqnarray} \label{cond5}
q^{\frac{2\chi}{\sqrt{1+\chi^2}}} = \frac{\sqrt{3(1+\chi^2)}-2+
    2\sqrt{1-3\chi^2}}{\sqrt{3(1+\chi^2)}+ 2-2\sqrt{1-3\chi^2}},
\end{eqnarray}
which comes from $R_q^2= r_q^2$.
This implies that the coordinate transformation~(\ref{boost}) is
pathological and  the Riemann curvatures are equivalent only
locally.

Since $\rho_e$ presents only for $\chi>0$, the local properties of the
solution with $\chi \leq 0$ are exactly equivalent to the static
hyper-cylindrical solution.
Now let us check the equivalence of the covariant derivatives of
the Riemann tensor between the two solutions.
For this purpose, we construct a Riemann normal coordinates around
the point ${\cal P}$.
In this coordinates the connection vanishes at the event ${\cal
P}$.
Therefore, the covariant derivatives of the Riemann tensor is
given by the simple derivative with respect to $\rho$.
Since the Riemann tensor in the frame is independent of $q$, their
derivatives become automatically independent of $q$.
In summary, the geometry of the new metric is locally equivalent to that
of the static hyper-cylindrical solution.
%

\section{Properties of the stationary solution}
Next, let us investigate the global property of the metric such as
the position of horizon. Consider the metric,
\begin{eqnarray}\label{ds2:3}
ds^{2}&=&g_{tt}(\rho) dt^2 + 2g_{tz}(\rho) dt dz +
g_{\rho\rho}(\rho) d\rho^2 + g_{zz}(\rho) dz^2 ,
\end{eqnarray}
where we have ignored $\theta$ and $\phi$ coordinates due to the
spherical symmetry. The Killing vectors for $t$- and
$z$-directions are $\displaystyle \xi_{(t)}\equiv
\left(\frac{\partial}{\partial t}\right)_{\rho,z}$ and
$\displaystyle \xi_{(z)}\equiv \left(\frac{\partial}{\partial
z}\right)_{t,\rho}$. The scalar products of these Killing vectors
generate the metric:
\begin{eqnarray} \label{..}
\xi_{(t)}\cdot \xi_{(t)} = g_{tt}, ~~~ \xi_{(t)}\cdot \xi_{(z)}=
g_{tz}, ~~~~\xi_{(z)}\cdot \xi_{(z)}= g_{zz}.
\end{eqnarray}
The Killing vector of a stationary observer moving with velocity
$\displaystyle v\equiv \frac{dz}{dt}$ relative to the asymptotic
rest frame is
\begin{eqnarray} \label{Kv:st}
\xi_{(obs)}= \frac{\xi_{(t)}+v \xi_{(z)}}{ |\xi_{(t)}+v
\xi_{(z)}|} = \frac{\xi_{(t)}+v \xi_{(z)}}{ (-g_{tt}-2v g_{tz}-v^2
g_{zz})^{1/2}} .
\end{eqnarray}
The stationary observer at a given $\rho$ cannot have arbitrary
velocity. Only the following value of $v$ are allowed, for which
the 4-velocity $\xi_{(obs)}$ lies inside the future light cone,
\begin{eqnarray*}
(\xi_{(t)}+v \xi_{(z)})^2= g_{tt}+2v g_{tz}+v^2 g_{zz}<0 .
\end{eqnarray*}
Thus, the velocity of stationary observers are constrained by,
\begin{eqnarray} \label{v:range}
&&v_{min} < v < v_{max}, ~~~~\mbox{for } g_{zz}>0,\\
&& v< v_{min}~~ \mbox{ or }~~ v> v_{max}, ~~~~\mbox{for } g_{zz}<0
,\nn
\end{eqnarray}
where
\begin{eqnarray} \label{ve}
v_{min}= \bar v-\sqrt{\bar v^2-g_{tt}/g_{zz}}, ~~~~ v_{max} =\bar
v+\sqrt{\bar v^2-g_{tt}/g_{zz}};~~~~ \bar v\equiv
-\frac{g_{tz}}{g_{zz}}=-\frac{FX}{H-FX^2} .
\end{eqnarray}
Far from the string, $v_{min}=-1$ and $v_{max}=1$ with $g_{zz}>0$. As
$\rho$ decreases, $v_{min}$ increases. $v_{min}$ becomes zero when
$\displaystyle \frac{g_{tt}}{g_{zz}}=0$, which is the static limit:
\begin{eqnarray} \label{ergo}
\rho_e = \frac{1+
q^{\sqrt{1+\chi^2}/(2\chi)}}{1-q^{\sqrt{1+\chi^2}/(2\chi)}}K.
\end{eqnarray}
The static limit exists only for $\chi> 0$ since we restrict
ourselves to the region with $\rho \geq K$. For $\chi<0$, $g_{zz}$
becomes negative for $\rho < \rho_t$ with
\begin{eqnarray} \label{rho:t}
 \rho_t=
\frac{1+q^{\sqrt{1+\chi^2}/(2|\chi|)}}{1-
q^{\sqrt{1+\chi^2}/(2|\chi|)}} K.
\end{eqnarray}
Therefore, the $z$ coordinates becomes timelike there. For the
time being, we consider the case $\chi>0$ and defer $\chi \leq 0$
case.

Inside the static limit $(\rho=\rho_e)$, all stationary observers must
move along the hyper-cylindrical solution with positive velocity.  As
$|q|$ increases, the position of the static limit, $\rho_e$, increases
so that the ergosphere encompasses the entire space at $q=1$. The
allowed range of the velocity narrows down until the limits $v_{min}$
and $v_{max}$ coalesce at $\rho_H=K$ which satisfies
\begin{eqnarray} \label{hor}
\sqrt{\bar v^2 - \frac {g_{tt}}{g_{zz}}} =
    \frac{1-q^2}{\left|1-q^2\left(\frac{1-K/\rho_H}{
            1+K/\rho_H}\right)^{\frac{4\chi}{\sqrt{1+\chi^2}}}
            \right|}\left(\frac{1-K/\rho_H}{
            1+K/\rho_H}\right)^{\frac{2\chi}{\sqrt{1+\chi^2}}}
            =0 .
\end{eqnarray}
In the comoving frame with velocity $\bar v$, the position of the
horizon is present when the metric $g'_{tt}$ in the comoving frame
vanishes:
\begin{eqnarray} \label{gtt'}
g'_{tt} = g_{tt}+2 \bar v g_{tz}+ \bar v^2 g_{zz}
    =-\frac{FH}{H-FX^2}
 = \frac{1-q^{2}}{
    1-q^{2}\left(\frac{1-K/\rho}{
 1+K/\rho}\right)^{\frac{4\chi}{\sqrt{1+\chi^2}}}
    }\left(\frac{1-K/\rho}{
            1+K/\rho}\right)^{\frac{2(\chi+1/\sqrt{3})}{
            \sqrt{1+\chi^2}}}=0
            .
\end{eqnarray}
The solution of Eq.~(\ref{gtt'}) exists for positive $\chi$ at
$\rho=\rho_H =K$. Therefore, the solution~(\ref{Sol}) describes a
hyper-cylindrical solution  without naked singularity. The velocity of
the horizon is given by
\begin{eqnarray} \label{vH}
v_H = -\left.\frac{g_{tz}}{g_{zz}}\right|_{\rho=K}= q , ~~~~0
~~\mbox{ for } \chi=0 .
\end{eqnarray}

Let us now analyze for the case $\chi <0$. Since the
$z$-coordinate becomes time-like for $\rho<\rho_t$, the analysis
for a stationary observer is not appropriate to probe an event
horizon. Consider the coordinate transformation from
$(t,\rho,z)\rightarrow (\tau,\rho, z)$ with $\tau=v t+z $, where
$v$ is a constant velocity. The metric~(\ref{ds2:3}) now becomes
\begin{eqnarray} \label{ds2:31}
ds^2= \frac{g_{tt}}{v^2}d\tau^2+\frac{2}{v}\left(g_{tz}-
\frac{g_{tt}}{v}\right)d\tau dz+\left(\frac{g_{tt}}{v^2}-
    \frac{2g_{tz}}{v}+g_{zz}\right)dz^2 + g_{\rho\rho}
    d\rho^2 .
\end{eqnarray}
At a point ${\cal Q}=(\tau,\rho, z)$, we may find a locally
orthogonal metric by setting the velocity $\dsp v=
\frac{g_{tt}(\rho)}{g_{tz}(\rho)}$. Now, the metric at the point
${\cal Q}$ becomes
\begin{eqnarray} \label{ds2:Q}
\left.ds^2\right|_{Q} =
\frac{g_{tz}^2(\rho)}{g_{tt}(\rho)}d\tau^2+
\left(g_{zz}(\rho)-\frac{g_{tz}^2(\rho)}{g_{tt}(\rho)}
    \right)dz^2 + g_{\rho\rho}(\rho) d\rho^2 .
\end{eqnarray}
Note that the $zz$ components of the metric at ${\cal Q}$,
\begin{eqnarray} \label{gzz'}
g_{zz}^{\cal Q}\equiv
g_{zz}(\rho)-\frac{g_{tz}^2(\rho)}{g_{tt}(\rho)}=H =
    \frac{(1-q^2)\left|\frac{1-K/\rho}{1 +
K/\rho}\right|^{\frac{2(|\chi|+1/\sqrt{3})}
    {\sqrt{1+\chi^2}}}}{
    1-q^2\left|\frac{1-K/\rho}{1 +
K/\rho}\right|^{\frac{4|\chi|}
    {\sqrt{1+\chi^2}}}}\, , \nn
\end{eqnarray}
is positive definite for all ${\cal Q}$ if $\chi \leq 0$. This
value vanishes at point ${\cal Q}_K=(\tau,K,z)$. We may write the
geodesic equation in this frame of reference. For fixed $z$, the
radially outgoing light-like geodesic satisfies
\begin{eqnarray} \label{lgeodsic}
\left.\frac{d\rho}{d\tau}\right|_{\cal Q}=
\sqrt{-\frac{g_{tz}^2(\rho)}{g_{tt}(\rho) g_{\rho\rho}(\rho)}} =
\sqrt{\frac{q^2}{1-q^2}\frac{\left(
    1-\left|\frac{1-K/\rho}{1 +
K/\rho}\right|^{\frac{4|\chi|}
    {\sqrt{1+\chi^2}}}\right)^2}{1-q^2
    \left|\frac{1-K/\rho}{1 +
    K/\rho}\right|^{\frac{4|\chi|}
    {\sqrt{1+\chi^2}}} }} \left(1+\frac{K}{\rho}\right)^{-1-
        \frac{\sqrt{3}+\chi}{\sqrt{1+\chi^2}}}
    \,\left(1-\frac{K}{\rho}\right)^{-1+
        \frac{\sqrt{3}+\chi}{\sqrt{1+\chi^2}}}
  .
\end{eqnarray}
The term inside the square root does not vanish for $\rho \geq K$
and $q\neq 0$. At point ${\cal Q}_K$, the
velocity~(\ref{lgeodsic}) vanishes only for $\chi
>-1/\sqrt{3}$. Therefore, the event horizon forms for
$\chi>-1/\sqrt{3}$ at $\rho=K$. This result is exactly the same as that
of the static hyper-cylindrical solution~(\ref{lee}). With this
analysis, we may argue that the structure of the event horizon for both
metrics are the same. Therefore, the solution~(\ref{Sol}) with  $(K,\chi
\leq 0,q)$ are equivalent to that with $(K,\chi,0)$ up to coordinate
transformation.

\section{Summary and discussions}
We have obtained a general stationary hyper-cylindrical solution with
arbitrary tension and momentum along the fifth coordinate which has
non-trivial $g_{04}$ component decreasing as $1/\rho$ for large radial
distance $\rho$. The solution shows an exact local equivalence to the
static hyper-cylindrical solution, but the solution with $\chi>0$ is
globally different from the static one except for the parameters
satisfying
Eq.~(\ref{cond5}).

We briefly have analyzed some physical properties of the metric
such as an event horizon and a static limit. Let us briefly
discuss the area of the horizon. The 2-dimensional surface area
for a fixed $\rho$, $z$, and $t$ can be obtained by integrating
the $\theta$ and $\phi$ coordinates,
\begin{eqnarray} \label{area:3}
S(\rho) = 4\pi \rho^2 G(\rho) =4 \pi
\rho^2\left|1-\frac{K}{\rho}\right|^{2-
    \frac{4}{\sqrt{3(1+\chi^2)}} }
    \left|1+\frac{K}{\rho}\right|^{2+
    \frac{4}{\sqrt{3(1+\chi^2)}} }.
\end{eqnarray}
The area for a given $\rho$ diverges for $\displaystyle
|\chi|<\frac{1}{\sqrt{3}}$ and vanishes for $\displaystyle
|\chi|>\frac{1}{\sqrt{3}}$ as $\rho \rightarrow K$. Only at
$\displaystyle |\chi|=\frac{1}{\sqrt{3}}$, a finite horizon area
is obtained:
\begin{eqnarray} \label{Area}
 S =64\pi K^2\, .
\end{eqnarray}

We consider only for $\chi\geq 0$. The distance $|\delta z(\rho)|$
corresponding to a unit length in $z$-direction for an asymptotic
observer in a constant $t$ surface is measured by
\begin{eqnarray}\label{dist:z}
|\delta z|_t^2 = g_{zz} =
 \frac{ 1-q^2\left|\frac{1-K/\rho}{1+K/\rho}\right|^{\frac{4
    \chi}{\sqrt{1+
    \chi^2} }}}{1-q^2}
    \left|\frac{1-K/\rho}{1+K/\rho}\right|^{\frac{2
    (1/\sqrt{3}-\chi)}{\sqrt{1+
    \chi^2} }} .
\end{eqnarray}
This behaves as
\begin{eqnarray} \label{..}
\lim_{\rho\rightarrow K}|\delta z|_t^2 =
 \frac{1}{1-q^2}
    \left|\frac{1-K/\rho}{1+K/\rho}\right|^{\frac{2
    (1/\sqrt{3}-\chi)}{\sqrt{1+
    \chi^2} }} .
\end{eqnarray}
Therefore,  $\delta z$  at the horizon vanishes for
$\chi<1/\sqrt{3}$, and  diverges for $\chi>1/\sqrt{3}$. Only at
$\chi = 1/\sqrt{3}$ a finite measure along $z$-coordinate at the
horizon are defined.

The surface area of the hyper-cylindrical solution for unit length of
$z$ with respect to an asymptotic observer is given by multiplying
$S(\rho)$ and $\delta z$:
\begin{eqnarray} \label{r}
A = S(\rho) |\delta z|=4\pi \rho^2 \left(\frac{
1-q^2\left|\frac{1-K/\rho}{1+K/\rho}\right|^{\frac{4
    \chi}{\sqrt{1+
    \chi^2} }}}{1-q^2}\right)^{1/2}
    \left|1-\frac{K}{\rho}\right|^{2-
    \frac{\sqrt{3}+\chi}{\sqrt{1+\chi^2}} }
    \left|1+\frac{K}{\rho}\right|^{2+
    \frac{\sqrt{3}+\chi}{\sqrt{1+\chi^2}} } .
\end{eqnarray}
The exponent $2-(\sqrt{3}+\chi)/\sqrt{1+\chi^2}$ is a non-negative
function which vanishes only at $\chi=1/\sqrt{3}$. Therefore, The
area at the horizon $\rho=K$ vanishes for $0\leq \chi \neq
1/\sqrt{3}$. For $\chi = 1/\sqrt{3}$, the area becomes $64\pi K^2
/(1-q^2)$. Since the area of the horizon is proportional to the
entropy of the black string, the hyper-cylindrical solution with $
\chi\neq 1/ \sqrt{3}$ has zero entropy. In the stability context,
this implies the hyper-cylindrical solution will be unstable.

In this paper, we write the surface $\rho=K$ as an event horizon
for $\chi=1/\sqrt{3}$ and a naked singularity for $|\chi| \neq
1/\sqrt{3}$. However, the exact identity of the surface $\rho=K$
is ambiguous yet and is under investigation by Kang et.
al.~\cite{kang}. As we have summarized in Sec. II, the properties
of the metric at $\rho=K$ becomes singular for $|\chi|\neq
1/\sqrt{3}$. Because the singularity is located at the same
position as the horizon, its physical properties are different
from the usual case.

The present metric provides an interesting test ground of the
Mach's principle (Inertia generates gravity) for linear motions.
In a literal sense, the linear motion should generate a velocity
frame dragging effect if the Mach's principle is correct.
According to the present result of the metric, the linear motion
along the fifth coordinate does not develop a local difference of
spacetime. However, it changes the global structures of spacetime
for some parameter range. Since the presence of frame dragging by
inertial motion is a controversial subject today, this result
provides an interesting inspection.
\begin{acknowledgments}
We are very grateful to Gungwon Kang for helpful discussions. This
work was supported by Daejin University Research Grants in 2006.
\end{acknowledgments} \vspace{3cm}

\begin{appendix}

\section{Lorentz boost of the Riemann tensor an an event ${\cal P}$}
The non-vanishing components of Riemann tensor in the 1-form
basis~(\ref{coor}) at the event ${\cal P}$ become
\begin{eqnarray} \label{curvatures}
R_{0101}&=&R_1+R_q, ~~~~ R_{0114} = R_{1401}=r_q, ~~~~
R_{1414} =-R_1+R_q, \nn\\
R_{0202}&=&R_{0303}=R_2 -\frac{1}{2}R_{q}, ~~~~ R_{0224}
=R_{2402}=R_{0334}=R_{3403}= -\frac{r_q}{2} ,\nn \\ ~~~~
 R_{2424}&=&R_{3434}=-R_2-\frac{1}{2}R_q, ~~~~
R_{0404} =
    \frac{1-3\chi^2}{1+\chi^2}\frac{4\kappa^2}{3
    \rho^4(1-K^2/\rho^2)^4}\left|\frac{1+K/\rho}{1 -
    K/\rho}\right|^{-\frac{4}
    {\sqrt{3}\sqrt{1+\chi^2}}},
 \nn \\
R_{1212} &=&R_{1313}= - \frac{4}{\sqrt{3(1+\chi^2)}}\frac{\kappa
(1-
    \sqrt{3(1+\chi^2)}K/\rho+K^2/\rho^2)}{
    \rho^3 (1-K^2/\rho^2)^4}\left|\frac{1+K/\rho}{1 -
K/\rho}\right|^{-\frac{4}
    {\sqrt{3}\sqrt{1+\chi^2}}},\nn \\
R_{2323} &=&\frac{4}
    {3(1+\chi^2)} \frac{K(2 K/\rho-\sqrt{3(1+\chi^2)}\,)
        (\sqrt{3(1+\chi^2)}\,K/\rho -2)}{\rho^3
(1-K^2/\rho^2)^4}\left|\frac{1+K/\rho}{1 -
K/\rho}\right|^{-\frac{4}
    {\sqrt{3}\sqrt{1+\chi^2}}}, \nn
\end{eqnarray}
where
\begin{eqnarray} \label{R:corr}
R_1 &=&\frac{4K\chi}{\sqrt{1+\chi^2}}\frac{
    \left|\frac{1+K/\rho}{1 -
    K/\rho}\right|^{-\frac{4}
    {\sqrt{3}\sqrt{1+\chi^2}}}}{\rho^3(1-K^2/\rho^2)^4}
   \left[ \frac{4K}{\rho}-\frac{4}{\sqrt{3(1+\chi^2)}}
   \left(\frac{K^2}{\rho^2}
        +1\right)\right] ,\\
R_2&=&\frac{2K}{\sqrt{3(1+\chi^2)}} \frac{ \left|\frac{1+K/\rho}{1
-
    K/\rho}\right|^{-\frac{4}
    {\sqrt{3}\sqrt{1+\chi^2}}}}{
        \rho^3(1-K^2/\rho^2)^4}
    \left(1-\frac{4}{\sqrt{3(1+\chi^2)}}
    \frac{K}\rho+\frac{K^2}{\rho^2}\right) ,
     \nn \\
R_q &=&-\frac{4K\chi}{\sqrt{1+\chi^2}}\frac{
\left|\frac{1+K/\rho}{1 -
    K/\rho}\right|^{-\frac{4}
    {\sqrt{3}\sqrt{1+\chi^2}}}}{\rho^3(1-K^2/\rho^2)^4}
     \left(1-\frac{4}{\sqrt{3(1+\chi^2)}}
     \frac{K}\rho+\frac{K^2}{\rho^2}\right)
     \frac{1+q^2\left|\frac{1+K/\rho}{1 -
K/\rho}\right|^{\frac{4\chi}
    {\sqrt{1+\chi^2}}}}{1-q^2\left|\frac{1+K/\rho}{1 -
K/\rho}\right|^{\frac{4\chi}
    {\sqrt{1+\chi^2}}}},
   \nn \\
r_q &=& \frac{8K\chi}{\sqrt{1+\chi^2}}\frac{
\left|\frac{1+K/\rho}{1 -
    K/\rho}\right|^{-\frac{4}
    {\sqrt{3}\sqrt{1+\chi^2}}}}{
    \rho ^3(1 -K^2/\rho^2)^4
    }\left(1-\frac{4}{\sqrt{3(1+\chi^2)}}
    \frac{K}\rho+\frac{K^2}{\rho^2}
   \right) \frac{q \left|\frac{1+K/\rho}{1 -
K/\rho}\right|^{\frac{2\chi}
    {\sqrt{1+\chi^2}}}}{1-q^2\left|\frac{1+K/\rho}{1 -
K/\rho}\right|^{\frac{4\chi}
    {\sqrt{1+\chi^2}}}}\nn  .
\end{eqnarray}
Now consider a Lorentz boost along the $z-$direction at the event
${\cal P}$:
\begin{eqnarray} \label{boost3}
\left(\begin{tabular}{c}
    ${x'}^0$\\ ${x'}^4$ \\\end{tabular}\right)
    &=&\left(\begin{tabular}{c}
    $\cosh \theta~~ \sinh\theta$\\ $\sinh\theta~~
        \cosh\theta$ \\ \end{tabular}\right)
\left(\begin{tabular}{c}
    $x^0$\\ $x^4$ \\\end{tabular}\right)
\end{eqnarray}
where $x^i$ is a coordinate system in the orthonormal frame at
${\cal P}$. In the boosted coordinates ${x'}^i$, the components of
Riemann tensor transform to be
\begin{eqnarray} \label{R:new}
R'_{0101} &=& R_1+ (\cosh^2\theta+\sinh^2\theta
    )R_q-2\cosh \theta \sinh\theta \, r_q, \\
R'_{1414} &=& -R_1+ (\cosh^2\theta+\sinh^2\theta)R_q
    -2\cosh\theta\sinh\theta\, r_q, \nn\\
R'_{0114} &=& (\cosh^2\theta+\sinh^2\theta)r_q
    - 2\cosh\theta\sinh\theta ~R_q, \nn \\
R'_{0202} &=& R_2-\frac{1}{2}\left[(\cosh^2\theta+\sinh^2\theta)
R_q-2
    \cosh \theta \sinh\theta\, r_q\right]
,\nn\\
R'_{0224} &=& -\frac{1}{2}\left[(\cosh^2\theta+\sinh^2\theta) r_q-
    2\cosh \theta\sinh \theta~R_q\right], \nn \\
R'_{2424} &=&
-R_2-\frac{1}{2}\left[(\cosh^2\theta+\sinh^2\theta)r_q-
    2\cosh \theta\sinh \theta~R_q\right], \nn \\
R'_{0404} &=&  R_{0404} .\nn
\end{eqnarray}
By choosing the boost parameter to satisfy
\begin{eqnarray*}
\tanh\theta+\frac{1}{\tanh\theta}= \frac{2R_q}{r_q}=
    \frac1{q}\left|\frac{1+K/\rho}{1 -
K/\rho}\right|^{-\frac{2\chi}
    {\sqrt{1+\chi^2}}}+ q\left|\frac{1+K/\rho}{1 -
K/\rho}\right|^{\frac{2\chi}
    {\sqrt{1+\chi^2}}},
\end{eqnarray*}
we have $R'_{0114}=0=R'_{0224}$. The unique solution of the
equation is
\begin{eqnarray} \label{theta:cond}
\tanh \theta =\left[ q\left|\frac{1+K/\rho}{1-K/\rho}
            \right|^{\frac{2\chi}{\sqrt{1+\chi^2}}}\right]^{
                \mbox{sign}(\rho-\rho_e)} .
\end{eqnarray}
where $\rho_e$ is the position of the ergosphere of the
metric~(\ref{Sol}),
\begin{eqnarray} \label{rhoe}
\rho_e= \frac{1+ |q|^\frac{\sqrt{1+\chi^2}}{2\chi}}{
    1- |q|^\frac{\sqrt{1+\chi^2}}{2\chi}} K .
\end{eqnarray}
The $q-$dependent part of components of Riemann tensor in
Eq.~(\ref{R:new}) is of the same form:
\begin{eqnarray} \label{R5}
&&(\cosh^2\theta+\sinh^2\theta)R_q-2\cosh \theta
    \sinh\theta \,r_q = -\mbox{sign}(\rho-\rho_e)\sqrt{R_q^2-r_q^2} \\
&&~~~~~~~~~=\mbox{sign}(\rho-\rho_e)
\frac{4K\chi}{\sqrt{1+\chi^2}}\frac{\left|\frac{1+K/\rho}{1
    -K/\rho}\right|^{-\frac{4}
    {\sqrt{3}\sqrt{1+\chi^2}}}
}{\rho^3(1-K^2/\rho^2)^4}
     \left|1-\frac{4}{\sqrt{3(1+\chi^2)}}
     \frac{K}\rho+\frac{K^2}{\rho^2}\right| ,\nn
\end{eqnarray}
where the last expression of Eq.~(\ref{R5}) is independent of $q$.
Therefore, the components of Riemann tensor~(\ref{R:new}) at the
event ${\cal P}$ in the transformed coordinate system becomes
independent of $q$.
%

\end{appendix}


\vspace{4cm}

\end{document}